\documentclass[pdflatex,sn-mathphys-num]{sn-jnl}

\usepackage{graphicx}%
\usepackage{multirow}%
\usepackage{amsmath,amssymb,amsfonts}%
\usepackage{amsthm}%
\usepackage{mathrsfs}%
\usepackage[title]{appendix}%
\usepackage{xcolor}%
\usepackage{textcomp}%
\usepackage{manyfoot}%
\usepackage{booktabs}%
\usepackage{algorithm}%
\usepackage{algorithmicx}%
\usepackage{algpseudocode}%
\usepackage{listings}%

\begin{document}

\title[Article Title]{Topological dressing method
for the Einstein-Maxwell equations
}


\author*{\fnm{Juri} \sur{Dimaschko}}



\affil{\orgname{Technische Hochschule Lübeck},

\orgaddress{\street{Max-Wartemann-Str. 18}, 

\city{Lübeck}, \postcode{23564}, \country{Germany}}}




\abstract{A regular method is proposed that makes it possible to obtain a new exact solution with a wormhole from any topologically trivial exact solution of the Einstein-Maxwell equations in an electrovacuum (topological dressing method). This solution has a structure similar to a thin-shell wormhole, but unlike it, it is exact and therefore does not require the presence of any other field sources. It is shown that the wormhole itself creates both gravitational and electromagnetic fields. The corresponding effective mass and effective charge are distributed over the surface of its throat and around it. Topological dressing of the Reissner-Nordström solution with zero effective mass and non-zero effective charge gives a new solution describing a traversable wormhole. It is shown that this solution is stable in the presence of external pressure.}

\keywords{Traversable wormhole, Einstein-Maxwell equations, electrovacuum}



\maketitle

\section{Introduction}\label{sec1}

With the advent of the wormhole concept in general relativity, it is constantly accompanied by the problem of traversability \cite{morris}. A wormhole provides a formal opportunity to move into “parallel space” - however, in the vast majority of cases this turns out to be physically impossible. The reason for the non-traversability is that the transition surface geometrically coincides with the event horizon. This means that from the point of view of an outside observer, crossing the transition surface takes an infinite time. This is exactly how the wormhole is constructed in the case of the Einstein-Rosen “bridge” \cite{einstein}: its throat coincides with the gravitational event horizon of the Schwarzschild solution. And that is why this wormhole is non-traversable.

For a spherical wormhole to be traversable, its radius must be greater than the gravitational one \cite{matt_visser}. However, various attempts to construct such a state either lead to the presence of so-called exotic matter, which has a negative mass density, or go beyond the scope of traditional GTR \cite{visser, poisson,sharif,zubair,khusnutdinov}. To date, within the framework of GTR, the only regular example of a traversable wormhole is known, which is not based on the use of exotic matter. This is a Bronnikov-Ellis wormhole \cite{bronnikov,ellis}, in which the function of matter is performed by a massless scalar field.

Interest in wormholes has long been associated mainly with the theoretical possibility of influencing the topology of space and using this effect for interstellar travel \cite{morris}. However, the results obtained over the past 10 years on the detection of gravitational waves from binary black holes \cite{abbott_1,abbott_2}, as well as the optical observation of the shadow from a supermassive black hole \cite{collaboration_eth}, have given a new concrete impetus to a more detailed study of the possible structure of wormholes. In the case of shadow effects, we are talking about observing the static structure of curved space near a wormhole \cite{oghami}, and in the case of gravitational waves, about the dynamics of oscillations (quasi-normal oscillatory modes \cite{zhao,manfredia}). Overall, this opens up new opportunities for testing the principles of general relativity and, if necessary, refining them.

The present paper proposes a general method that allows one to obtain a new solution with a wormhole from an arbitrary topologically trivial solution of the Einstein-Maxwell equations. The method is based on the covariance of the Einstein-Maxwell equations with respect to coordinate transformations and has a form of the parametric generalization of the Einstein-Rosen transformation. In particular, this approach makes it possible to obtain solutions that describe a traversable wormhole, remaining entirely within the framework of the Einstein-Maxwell equations and without resorting to exotic matter.

Below a system of units is used in which the speed of light and the gravitational constant are equal to one (\(c=G=1\)), as well as the metric signature (\(+,-,-,-\)).

\section{Topological dressing method}\label{sec2}
\subsection{Transformation of space topology}

We will proceed from the usual expression for the action considering the possible presence of an electromagnetic field \cite{landau} 
\begin{equation}
S=\frac{1}{16\pi}\int \left ( R - F_{\mu \nu } F^{\mu \nu } \right ) \sqrt{-g} d^{4}x.
\end{equation} Here \(R\) is the scalar curvature of space-time, \(g=\det \left ( g_{\mu \nu } \right )\) is the determinant of the metric tensor, and \( F_ {\mu \nu} \) is the electromagnetic field tensor. 

All local geometric properties of space are specified by the metric tensor \(g_ {\mu \nu} \) and the principle of least action
\begin{equation}
\frac{\delta S}{\delta g_{\mu \nu }}=0,
\end{equation} which leads to the Einstein-Maxwell equations
\begin{equation}
R_{\mu \nu }-\frac{1}{2}g_{\mu \nu }R=8\pi T_{\mu \nu }.
\end{equation} Here the stress-energy tensor \(T_{\mu \nu }\) is expressed through the electromagnetic field tensor \(T_{\mu \nu }\) and has the form
\begin{equation}
T_{\mu \nu }=\frac{1}{4\pi }\left (g^{\alpha \beta }F_{\mu \alpha }F_{\nu \beta  }-\frac{1}{4}g_{\mu \nu }F_{\alpha  \beta  }F^{\alpha  \beta  }  \right ).
\end{equation} The metric tensor \(g_{\mu \nu }\)  and the electromagnetic field tensor \(F_{\mu \nu }\) are defined on a certain coordinate space \(\mathcal{M}\), which we will call the basis manifold. The physical state of the system is completely determined by the pair \((\mathcal{M},g_{\mu \nu } )\).

\begin{figure*}
  \centering
  \includegraphics[width=11cm,height=4cm]{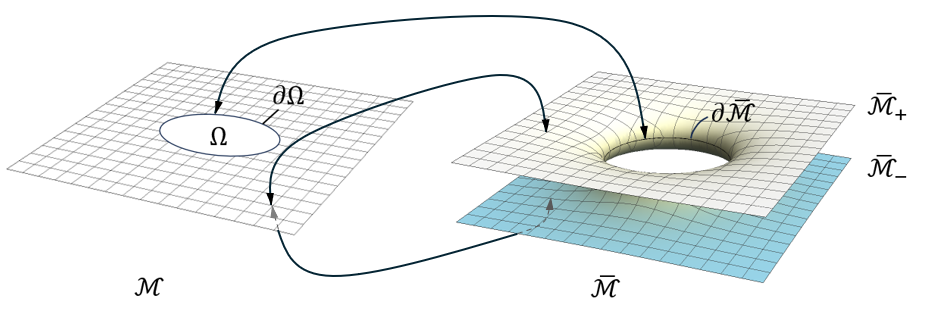}
  \caption{Two-sheeted transformation \(x \rightarrow \bar{x}\) of the original one-sheet manifold \(\mathcal{M}\) into a new two-sheeted manifold \(\bar{\mathcal{M}}\). }
\end{figure*}

We will start from some known solution to the Einstein-Maxwell equations (3,4), represented by metric tensor \(g_{\mu \nu }\). It is assumed that the solution is defined on some (also known) basis manifold \(\mathcal{M}\), topologically equivalent to the Minkowski space. We will call such a basis manifold \(\mathcal{M}\) and any corresponding physical state \((\mathcal{M},g_{\mu \nu } )\) one-sheeted.

The goal of the further construction is to obtain from the one-sheeted state \((\mathcal{M},g_{\mu \nu } )\), which has the trivial topology of Minkowski space, a new physical state \((\bar{\mathcal{M}},\bar{g}_{\mu \nu } )\), which is defined on a new manifold \(\bar{\mathcal{M}}\) having the topology of a wormhole. We call such a procedure topological dressing of the original one-sheet state.

To implement this procedure, we turn to the well-known topological construction of Einstein-Rosen \cite{einstein}, when after a two-sheet coordinate transformation, the one-sheet Schwarzschild solution turns into a wormhole connecting two asymptotically flat sheets. We construct this procedure as a direct generalization of the Einstein-Rosen transform.

To do this, we choose some general compact region  \(\Omega \subset \mathcal{M} \) with boundary  \(\partial \Omega\) in the original manifold \(\mathcal{M}\) as well as some smooth two-sheet function\( f(\bar{x}) \), defined on the new manifold \(\bar{\mathcal{M}}\). It is assumed that the new manifold \textbf{\(\bar{\mathcal{M}}\)} can be represented as a sum of three disjoint regions
\begin{equation}
\bar{\mathcal{M}}=\bar{\mathcal{M}}_{+} \cup \partial \bar{\mathcal{M}} \cup \bar{\mathcal{M}}_{-} 
\end{equation}in such a way that the function \( f(\bar{x}) \) generates three one-to-one smooth mappings (see Fig.1)
\begin{equation}
\bar{\mathcal{M}}_{+} \leftrightarrow \mathcal{M} \setminus \Omega ,
\end{equation}
\begin{equation}
\bar{\mathcal{M}}_{-} \leftrightarrow \mathcal{M} \setminus \Omega ,
\end{equation}
\begin{equation}
\partial \bar{\mathcal{M}} \leftrightarrow \partial \Omega . 
\end{equation}

In what follows, we will call the new manifold \(\bar{\mathcal{M}}\) two-sheeted, and its parts \(\bar{\mathcal{M}}_{+},\bar{\mathcal{M}}_{-}\) and \(\partial \bar{\mathcal{M}}\) , respectively, first sheet, second sheet and throat.

Let's consider the coordinate transformation specified by the function \( f(\bar{x}): x=f(\bar{x}) \). This transforms the original one-sheet physical state  \((\mathcal{M},g_{\mu \nu } )\) into a new two-sheeted state \((\bar{\mathcal{M}},\bar{g}_{\mu \nu } )\), where
\begin{equation}
\bar{g}_{\mu \nu }=\frac{\partial x^{\alpha }}{\partial \bar{x}^{\mu }}\frac{\partial x^{\beta }}{\partial \bar{x}^{\nu  }}g_{\alpha \beta } .
\end{equation}
Due to the covariance of the GTR equations with respect to coordinate transformations, the new metric tensor \(\bar{g}_{\mu \nu }\) is also a solution to the Einstein-Maxwell equations. This means that the new two-sheeted state \((\bar{\mathcal{M}},\bar{g}_{\mu \nu } )\) obtained in this way is physical - just like the original one-sheet state \((\mathcal{M},g_{\mu \nu } )\).

Consider, further, the Jacobian of the coordinate transformation \(x \rightarrow \bar{x}\):
\begin{equation}
J=\det \left (\frac{\partial x^{\mu }}{\partial \bar{x}^{\nu }}  \right ).
\end{equation}
According to conditions (6-8) it is not zero on both sheets  \(\bar{\mathcal{M}}_{+},\bar{\mathcal{M}}_{-}\)  and vanishes at the throat: \(J(\bar{x})=0 \) for \(\bar{x} \in \partial \bar{\mathcal{M}}\). Due to the fact that when transforming the coordinates  \(x \rightarrow \bar{x}\), the determinant \(g\) of the metric tensor is also transformed according to the relation \( g\rightarrow \bar{g} = J^2 g\), the new metric \(\bar{g}_{\mu \nu }\) is degenerated at the throat \(\partial \bar{\mathcal{M}}\):
\begin{equation}
\bar{g}=0 \quad \quad  (\bar{x} \in \partial \bar{\mathcal{M}}).
\end{equation}
Note that the condition for the degeneracy of the metric \(\bar{g}=0\), which holds at the throat \(\partial \bar{\mathcal{M}}\), is weaker than the condition for the event horizon, which is \(\bar{g}_{tt}=0\). This means that in the case of a diagonal form of the metric, any event horizon is a throat, but not every throat is an event horizon.

The physical state arising as a result of the described topological dressing procedure is defined on the two-sheeted manifold \(\bar{\mathcal{M}}\) and has the topology of a wormhole. As already mentioned, the topological dressing procedure itself is a generalization of the Einstein-Rosen transformation in the following sense:

A) the coordinate transformation function \(x=f(\bar{x})\) is an arbitrary two-sheeted function that satisfies conditions (6–8);

B) the excluded region \(\Omega\) is also arbitrary.

On the other hand, the Einstein-Rosen transform for the Schwarzschild solution is a special case of topological dressing in the following sense:

A) the coordinate transformation function \(r=\bar{r}^2+2M\) affects only the radial coordinate and is only quadratic;

 B) the region \(\Omega\) is fixed and as a sphere \(r<2M\); its radius is also fixed and coincides with the gravitational radius.

The illustration presented in Fig.1 in relation to the Einstein-Rosen transform is well known and is given in many works as a topological image of a wormhole.

\subsection{Topology as a dynamic variable}

The topological dressing procedure guarantees the existence of a two-sheeted physical state, which corresponds to a new solution \(\bar{g}_{\mu \nu }\) of the Einstein-Maxwell equations. However, this procedure itself is largely arbitrary and does not make it possible to uniquely find a new two-sheeted manifold  \(\bar{\mathcal{M}}\), which determines the topology of the new physical state \((\bar{\mathcal{M}},\bar{g}_{\mu \nu } )\).

To jointly determine both the new metric  \(\bar{g}_{\mu \nu }\), which determines the local geometry, and the new basic manifold  \(\bar{\mathcal{M}}\),  which determines the new global topology of the space, it is necessary to apply the principle of least action, taking into account the complete dependence of the action \(S\) on both arguments:
\begin{equation}
\delta S = \frac{\delta S}{\delta \bar{g}_{\mu \nu }} \delta \bar{g}_{\mu \nu } + \frac{\delta S}{\delta \bar{\mathcal{M}}} \delta \bar{\mathcal{M}}=0.
\end{equation}
Due to the arbitrariness of the variations, the following equations arise: 
\begin{equation}
\frac{\delta S}{\delta \bar{g}_{\mu \nu }}=0,
\end{equation}
\begin{equation}
\frac{\delta S}{\delta \bar{\mathcal{M}}}=0.
\end{equation}
The first of them gives the Einstein-Maxwell equations that determine the new metric  \(\bar{g}_{\mu \nu }\). 

The second equation gives a new condition, additional to the Einstein-Maxwell equations. It determines a new basis manifold \(\bar{\mathcal{M}}\). Using this condition implies solving equations (14) with an arbitrary choice of a new basis manifold \(\bar{\mathcal{M}}\), determining the dependence of the value of action (1) on the choice of \(\bar{\mathcal{M}}\), \(S=S(\bar{\mathcal{M}})\), and then finding the minimum for this dependence. This is a very complex problem, but it allows for the following simplification.

Since, due to the presence of the  factor \(J=\sqrt{-g}\) in the integrand, integration in action (1) is invariant with respect to any subsequent one-sheet transformation of coordinates \(\bar{x} \rightarrow \bar{\bar{x}}\), the specific choice of the  \(f(\bar{x})\) function does not affect the magnitude of the action \(S\). The only property of the function \(f(\bar{x})\) which matters is its range of values \(\mathcal{M} \setminus \Omega\), determined by the choice of the excluded region \( \Omega \). Thus, in the variational equation (14) it is enough to vary only the region \( \Omega \)  (and hence its boundary \(\partial \Omega\)):
\begin{equation}
\frac{\delta S}{\delta \Omega}=0.
\end{equation}

The variational derivative \(\delta / \delta \Omega\) here has the meaning of a complete set of derivatives with respect to all parameters that determine the shape of the region \(\Omega\). For example, if we choose \( \Omega\) from a family of circles of radius \(a\), then the variational derivative \(\delta / \delta \Omega\) coincides with the usual derivative \(d/da\); if we choose \(\Omega\) from the family of ellipses with semi-axes \(a_1\) and \(a_2\), then \(\delta / \delta \Omega\) is a 2-gradient \((\partial/\partial a_1,\partial /\partial a_2)\), etc.

Further, in view of the same invariance of expression (1) for the action \(S\) with respect to any coordinate transformations, its value can be obtained in accordance with (6, 7) by passing from integration over the variable \(\bar{x} \) on the two-sheeted manifold \(\bar{\mathcal{M}}\) with a new metric \(\bar{g}_{\mu \nu }\) - to the double integration over the variable \(x\) on the exterior of the region  \( \Omega \) in the one-sheet manifold \(\mathcal{M}\) using the original (one-sheet) metric \(g_{\mu \nu }\):
\begin{equation}
\begin{split}
S(\Omega )=\frac{1}{16\pi}\int_{\bar{\mathcal{M}}} \left ( \bar{R} - \bar{F}_{\mu \nu } \bar{F}^{\mu \nu } \right ) \sqrt{-\bar{g}} d^{4}\bar{x} =\frac{2}{16\pi}\int_{\mathcal{M} \setminus \Omega } \left ( R - F_{\mu \nu } F^{\mu \nu } \right ) \sqrt{-g} d^{4}x.
\end{split}
\end{equation}Here the factor 2 arises because of the double integration over the region  \(\mathcal{M} \setminus \Omega\) of the original one-sheet space \(\mathcal{M}\).

Finally, in the case of an electrovacuum, the only stress-energy tensor is that of the electromagnetic field (4), the trace of which is zero, \(T=0\). Therefore, the trace of the Ricci tensor also vanishes, \(R=0\), and the action takes on a simple form

\begin{equation}
S(\Omega )=-\frac{2}{16\pi}\int_{\mathcal{M} \setminus \Omega} F_{\mu \nu } F^{\mu \nu }  \sqrt{-g} d^{4}x.
\end{equation}
This means that the dependence of the action \(S\) on the region \(\Omega\) can be determined already before moving to the two-sheeted manifold \(\bar{\mathcal{M}}\). Expression (17) gives the dependence \(S(\Omega) \) directly based on the original one-sheet solution of the Einstein-Maxwell equations.

Thus, the topological dressing procedure involves the following actions:

1) A solution of the Einstein-Maxwell equations \(g_ {\mu \nu}(x) \) is selected, corresponding to one-sheet physical state  \((\mathcal{M},g_{\mu \nu } )\).

2) The action \(S(\Omega )\) depending on the choice of the inverse image of the throat \(\partial \Omega\) is calculated according to (17).

3) By minimizing \(S(\Omega )\),  the region \(\Omega\) and its boundary \(\partial \Omega\) are determined.

4) A two-sheet function \( f(\bar{x}) \) is chosen that provides a degenerate coordinate transformation \(x \rightarrow \bar{x}\) on the boundary \(\partial \Omega\).

5) Applying this transformation to the original one-sheet state \((\mathcal{M},g_{\mu \nu } )\) turns it into a new two-sheeted state \((\bar{\mathcal{M}},\bar{g}_{\mu \nu } )\)  having a wormhole topology.

\begin{figure*}
  \centering
  \includegraphics[width=13cm,height=4.5cm]{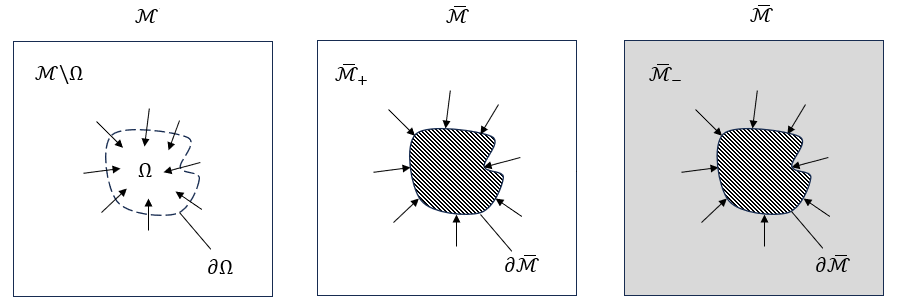}
  \caption{Distribution of the gravitational or electromagnetic field before and after the transformation of the original one-sheet space \(\mathcal{M}\) into a new two-sheeted space \(\bar{\mathcal{M}}\). }
\end{figure*}

As a result of topological dressing, a new physical state arises \((\bar{\mathcal{M}},\bar{g}_{\mu \nu } )\), in which the tensors  \(\bar{g}_{\mu \nu }\) and   \(\bar{F}_{\mu \nu }\) are defined on the entire two-sheeted manifold \(\bar{\mathcal{M}}\). Due to the fact that the transformation \(x \rightarrow \bar{x}\) according to (6, 7) establishes a one-to-one correspondence between each of the two sheets \(\bar{\mathcal{M}}_{+},\bar{\mathcal{M}}_{-}\) of the new two-sheeted space and the external region \(\mathcal{M} \setminus \Omega\)  of the original one-sheet space, the tensors \(\bar{g}_{\mu \nu }, \bar{F}_{\mu \nu } \) on both sheets coincide with those of the external region \(\mathcal{M} \setminus \Omega\) : \(g_{\mu \nu }=\bar{g}_{\mu \nu }, F_{\mu \nu }=\bar{F}_{\mu \nu }\). This means that after topological dressing, the field distribution on both the first and second sheets of the new two-sheeted space \(\bar{\mathcal{M}}\) are identical to the original field distribution in the part of the one-sheet space \(\mathcal{M}\) external to the region \(\Omega\) (see Fig. 2).

The resulting two-sheeted state  \((\bar{\mathcal{M}},\bar{g}_{\mu \nu } )\) is self-sufficient and has no external sources of the field, other than the gravitational and electromagnetic field itself. The only nonzero stress-energy tensor in the Einstein-Maxwell equations (3, 4) is that of the electromagnetic field. As will become clear from the two examples considered in the following, the natural region of localization of effective field sources in the two-sheeted space  \(\bar{\mathcal{M}}\) is the throat \(\partial \bar{\mathcal{M}}\) - the boundary between the first and second sheets - and its surrounding.

Note that the variational condition (15) either gives a stationary value of the throat size, or (if there is no stationary value) determines the direction of its change. In the first case, this fixes the stable state of the wormhole. In the second case, this is an indication of what external factors should be included in consideration to stabilize it. Then, the resulting parametric family of solutions can be considered as a variational ansatz.

\section{Two examples}\label{sec4}

Next, we look at two examples of using the topological dressing method.

\subsection{Schwarzschild wormhole}

The Schwarzschild solution describes a spherically symmetric state of space in the absence of an electromagnetic field. In spherical coordinates spherical coordinates \(\left ( r, \theta , \varphi  \right )\), defined on the one-sheet Minkowski space \(\mathcal{M}\), this solution is described by the metric

\begin{equation}
   \begin{split}
ds^{2}=\left (1-\frac{2M}{r} \right )dt^{2}-r^{2}d\Omega^{2} -\left (1-\frac{2M}{r}  \right )^{^{-1}}dr^{2}.
   \end{split}
\end{equation}
Here \(d\Omega ^{2}=d\theta ^{2}+\sin ^{2}\theta d\varphi ^{2}\) is the solid angle element, and the parameter \(M\) has the meaning of effective mass. This metric has a singularity at \(r=0 \) and an event horizon on a spherical surface \(r=2M\) (since \(g_{tt}=0\) on this surface).

Following the work of Einstein and Rosen \cite{einstein}, we first choose the spherical surface \(r=2M\) as the inverse image of the throat \(\partial \Omega\), and the mapping \(r=\bar{r}^2+2M\) as the two-sheeted function \(f(\bar{r})\). This mapping transforms the original one-sheet Minkowski space \(\mathcal{M}\) with coordinates \(\left (t, r, \theta , \varphi  \right )\) into a two-sheeted space \(\bar{\mathcal{M}}\) with coordinates \(\left (t, \bar{r}, \theta , \varphi  \right )\),  and the original metric (18), defined on \(\mathcal{M}\), into a new metric
\begin{equation}
   \begin{split}
ds^{2}=\frac{\bar{r}^{2}}{\bar{r}^{2}+2M }dt^{2} -\left (\bar{r}^{2}+2M  \right )^{2}d\Omega^{2} -4 \left (\bar{r}^{2}+2M   \right )dr^{2},
   \end{split}
\end{equation}
defined and non-singular on the entire two-sheeted space  \(\bar{\mathcal{M}}\). Here, in accordance with the general structure (5) of the two-sheet space  \(\bar{\mathcal{M}}\),  the first sheet  \(\bar{\mathcal{M}}_{+}\) is the region \(\bar{r}>0\), the second sheet  \(\bar{\mathcal{M}}_{-}\)is the region \(\bar{r}<0\),  and the throat \(\partial \bar{\mathcal{M}}\) is the spherical hypersurface \(\bar{r}=0\) connecting them.

Just like the original metric (18), the new metric (19) has an event horizon – a hypersurface  \(\bar{r}=0\). In this case, it coincides with the throat \(\partial \bar{\mathcal{M}}\), which makes the Einstein-Rosen wormhole non-traversable.

Note that this obstruction is not inevitable. If we choose as the inverse image of the throat \(\partial \Omega\) not a spherical surface \(r=2M\), but a parametric class of spherical surfaces \(r=a\), then the corresponding new metric has the form
\begin{equation}
   \begin{split}
ds^{2}=\left (1-\frac{2M}{\bar{r}^{2}+a } \right )dt^{2} -\left (\bar{r}^{2}+a \right )^{2}d\Omega^{2} - 
\left (1-\frac{2M}{\bar{r}^{2}+a } \right )^{-1} 4\bar{r}^{2}dr^{2}.
   \end{split}
\end{equation}
For \(a>2M\), this metric does not have an event horizon, since in this case the time component of the metric tensor is positive over the entire two-sheeted space, \(g_{tt}>0\).

In a vacuum, the trace of the curvature tensor is zero, \(R=0\). Therefore, the magnitude of action (1) is also zero. Since this magnitude does not depend on the radius a of the throat, \(S(a)=0\),  the value of radius \(a\) is indefinite and can take any positive value. At the same time, accounting for any additional stress-energy tensor allows us to use the class of metrics (20) as a variational ansatz that affects the dependence \(S(a)\)  and to determine the equilibrium radius \(a\) of the throat. If this equilibrium radius is in the region \(a>2M\), then the resulting wormhole is traversable. Otherwise (in particular, for \(a=2M\), as was taken by Einstein and Rosen), the metric (20) has an event horizon, which makes the wormhole non-traversable.

To determine the location of the sources of the gravitational field in the two-sheet solution (20), we find their distribution in the two-sheet space \(\bar{\mathcal{M}}\). As was stated earlier, in one-sheet \(r\)-coordinates it coincides with the field distribution on the one-sheet space \(\mathcal{M}\) in the region \(r>a\) and is expressed through the Christoffel symbol \( \Gamma_{tt}^r\):
\begin{equation}
     \mathfrak{g}_r = -\Gamma_{tt}^r = -\left(1 - \frac{2M}{r}\right) \frac{M}{r^2} \quad (r > a)
\end{equation}The corresponding field distribution on the first and second sheets is symmetrical with respect to the permutation of the sheets as is shown in  Fig.3.

\begin{figure}
  \centering
  \includegraphics[width=9cm,height=4.5cm]{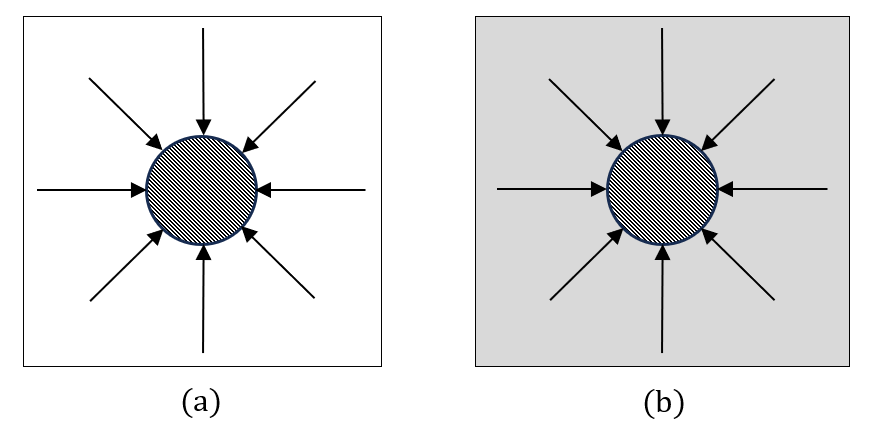}
  \caption{Identical distribution of the gravitational field around the throat (a) on the first sheet and (b) on the second sheet of the two-sheeted space \(\bar{\mathcal{M}}\). }
\end{figure}

In accordance with Gauss's theorem, this field distribution corresponds to the surface density of the effective mass at the throat of the wormhole 
\begin{equation}
    \sigma_M = \left(1 - \frac{2M}{a}\right) \frac{M}{4\pi a^2}
    \label{eq:sample_equation}
\end{equation}
as well as the volumetric density of the effective mass outside it
\begin{equation}
    \rho_M (r) = \frac{M^2}{2 \pi r^4} \quad (r > a).
\end{equation}
The total effective mass is
\begin{equation}
    4\pi a^2 \sigma_M + \int_a^\infty 4\pi r^2 \rho_M(r) dr = M
\end{equation}and coincides with the effective mass determined from the principle of correspondence with Newtonian  theory-

Relations (21–24) show that the two-sheeted Einstein-Rosen bridge metric (19), as well as its generalization (20), describe a self-consistent solution of the Einstein equations in vacuum. It is a source of gravitational field itself and does not require any external sources.

\subsection{Reissner-Nordström wormhole}

Next we take the massless Reissner-Nordström solution \cite{reissner} , which describes a spherically symmetric state of space, corresponding to zero effective mass \(M=0\)  and a non-zero effective charge \(Q \ne 0\). In spherical coordinates it is described by the metric
\begin{equation}
   \begin{split}
ds^{2}=\left (1+\frac{Q^{^2}}{r^{^2}}  \right )dt^{2}-r^{2}d\Omega^{2} -\left (1+\frac{Q^{^2}}{r^{^2}}  \right )^{^{-1}}dr^{2}.
   \end{split}
\end{equation}
This metric is defined on the one-sheet Minkowski space \(\mathcal{M}\) and has a singularity at \(r=0\). It has no event horizon, since at any point \(g_{tt}>0\). Metric (27) corresponds to two non-zero components of the electromagnetic field tensor \(F_{tr}=-F_{rt}=E_{r}\):
\begin{equation}
  E_{r}=\frac{Q}{r^{2}},
\end{equation}
which also indicates a singularity at \(r=0\).

Similarly to the work [2], we choose the spherical surface \(r=a\) as the inverse image of the throat \(\partial \Omega\) and the mapping \(r=\sqrt{\bar{r}^{2}+a^{2}}\) as the two-sheet function \(f(\bar{r})\). This mapping transforms the original one-sheet Minkowski space \(\mathcal{M}\) with coordinates \(\left (t, r, \theta , \varphi  \right )\) into a two-sheeted space \(\bar{\mathcal{M}}\) with coordinates \(\left (t, \bar{r}, \theta , \varphi  \right )\), and the original metric (25), defined on \(\mathcal{M}\), into a new metric

\begin{equation}
   \begin{split}
ds^{2}=\left (1+\frac{Q^{2}}{\bar{r}^{2}+a^2}  \right )dt^{2}-\left (\bar{r}^{2}+a^2  \right )d\Omega^{2}- \left (1+\frac{Q^{^2}+a^2}{\bar{r}^{2}}  \right )^{^{-1}}d\bar{r}^{2},
   \end{split}
\end{equation}
defined and non-singular on the entire two-sheeted space \(\bar{\mathcal{M}}\). This metric corresponds to the electric field distribution
\begin{equation}
E_{\bar{r}}=\frac{Q\bar{r}}{\left (\bar{r}^{2}+a^{2}  \right )^{3/2}},
\end{equation}
which is also non-singular.

Since the new metric (27) is defined in a connected double space \(\bar{\mathcal{M}}\) and has no event horizon, it describes the physical state of a space with a traversable wormhole. Since this wormhole has an effective mass of zero \(M=0\), we call it massless.

A massless wormhole is a source of electromagnetic field. In one-sheet \(r \) coordinates, its distribution coincides with the field distribution in the one-sheet space \(\mathcal{M}\) in the region \(r>a\), which has the Coulomb form (26). According to the Gauss theorem, this expression corresponds to the uniform distribution of the effective charge \(Q\) along the  throat \(r=a\), with the surface density of the effective charge
\begin{equation}
  \sigma_{Q}=\frac{Q}{4\pi a^{2}}.
\end{equation}
The total effective charge of a massless wormhole is
\begin{equation}
 4\pi a^{2} \sigma_{Q}= Q
\end{equation}
and coincides with the effective charge \(Q\), which follows from the principle of correspondence with Coulomb's law.

To determine the radius \(a\) of a massless wormhole, it is necessary to find the dependence of the total action (1) on this radius, \(S(a)\). As was shown above, this can already be done based on the original solution (25, 26), defined on the one-sheet Minkowski space. According to (17), the dependence of the action on the radius \(a\) has the form
\begin{equation}
S(a)= 2 \int_{r>a} \frac{E^{2}}{8\pi } d^{4}x = \frac{Q^{2}}{a} \Delta t,
\end{equation}
where \(\Delta t \) is the distance between two time hyperplanes. Since the dependence \(S(a)\) is monotonically decreasing, without taking into account factors external to the initial action (1), the massless wormhole itself is unstable. This factor counteracting the growth of radius \(a\) can be the pressure of the medium - for example, gas.

Due to the decrease in action with increasing radius \(a\), the system goes into a state where \(Q/a \ll 1\), the metric (25) tends to flatten, and the Newtonian approximation becomes applicable. In this approximation, the gas is not subject to the action of a nonuniform gravitational field, which allows us to consider the value of its pressure \(p\) as a constant independent of the coordinates. Taking into account uniform pressure, the expression for action (31) takes the form
\begin{equation}
S(a)= 2 \int_{r>a} \left (\frac{E^{2}}{8\pi }-p  \right ) d^{4}x 
\end{equation}
and leads to a nonmonotonic dependence \(S(a)\):
\begin{equation}
S(a)=\left (\frac{Q^2}{a}+\frac{8\pi a^3}{3}  \right )\Delta t.
\end{equation}
The function \(S(a)\) obtained in this way has a single minimum at the point
\begin{equation}
a=\left (\frac{Q^2}{8\pi p}  \right )^{1/4}
\end{equation}
corresponding to the equilibrium radius of a massless wormhole. Together with the two-sheet solution (27, 28), this determines the stable equilibrium state of a massless wormhole carrying an effective charge \(Q\) and stabilized by external pressure \(p\).

\section{Comparison with thin shells and RBH}
\subsection{Thin shells}

The two-sheet metric (20) of the Schwarzschild wormhole can be represented in one-sheet \(r\)-coordinates with the constraint \(r \ge a\):
\begin{equation}
   \begin{split}
ds^{2}=\left (1-\frac{2M}{r} \right )dt^{2}-r^{2}d\Omega^{2} -\left (1-\frac{2M}{r}  \right )^{^{-1}}dr^{2} \quad \quad (r \ge a).
   \end{split}
\end{equation}
This expression formally coincides with the Schwarzschild solution (18) in the region \(r \ge a\).

Similarly, the two-sheet metric (27) of the Reissner-Nordström wormhole can also be represented in the same one-sheet coordinates with the constraint \(r \ge a\):
\begin{equation}
   \begin{split}
ds^{2}=\left (1+\frac{Q^{^2}}{r^{^2}}  \right )dt^{2}-r^{2}d\Omega^{2} -\left (1+\frac{Q^{^2}}{r^{^2}}  \right )^{^{-1}}dr^{2} \quad \quad (r \ge a).
   \end{split}
\end{equation}
This expression also formally coincides with the massless Reissner-Nordström solution (27) in the region \(r \ge a\).

Both expressions are quite similar to those used within the two-sheet model of thin shells \cite{matt_visser}. The only (but decisive) difference is that in the thin shell model the domain of definition of the metric does not include the throat of the wormhole, i.e. the sphere \(r=a\). This circumstance forces us to separately consider the Einstein-Maxwell equations in the vicinity of the throat. In the thin shell model, this is done by calculating the jump in the Christoffel symbol at  \(r=a\), which leads to a \(\delta\)-shaped singularity of the Ricci tensor at the throat. In turn, this means the presence of matter with a \(\delta\)-shaped density at the throat. Further analysis clearly indicates that this matter has a negative mass density. This matter, which has not yet been discovered in nature, was called exotic.

Given the formal similarities between the topological dressing method proposed in this work and the thin shell model, it seems appropriate to directly compare these two approaches.

\underline {The thin shell model} is based on a specific solution ansatz, consisting of two one-sheet components. This ansatz is complemented by a special 'stitching' procedure treating these two components as two parts in a two-sheet state. The essential point here is the boundary condition and the corresponding limiting transition to an infinitely thin shell. Verification of the solution constructed in this way by direct substitution into the general-relativity equations beyond the framework of the same limiting passage is impossible.

\underline {The topological dressing method} is based on a fundamental property of the general relativity equations, their covariance with respect to coordinate transformations. This method allows us to directly obtain a new solution with a wormhole topology from a known solution of the general relativity equations, which has a trivial topology. Direct verification of the new solution obtained in this way does not present any difficulties.

Let us give a more detailed comparison of the two approaches using the Schwarzschild wormhole as an example. This is one of the first examples of the application of the thin shell method. It is described in detail in the well-known monograph \cite{matt_visser}, 15.2. Let us compare the physical description and main results of the thin shell method given in \cite{matt_visser} with the description and results of the topological dressing method.

A) Outside the throat (\(r>a\)), both approaches assume a vacuum, that is, a zero stress energy tensor. In both approaches, the metric outside the throat coincides with the Schwarzschild vacuum solution.

B) At the throat (\(r=a\)), \underline the thin shell model requires a density of matter of \(\delta\) shape. \underline {The topological dressing method} does not cause the appearance of matter in the throat, that is, the stress-energy tensor remains zero - just like the original Schwarzschild vacuum solution.

C) The source of the gravitational field in the two compared approaches is fundamentally different. \underline {In the thin shell model}, the source of the gravitational field is matter (usually exotic). \underline {In the method of topological dressing}, the source of the field is the curved space itself. The corresponding effective mass distribution includes both the surface density (22) at the throat and the volumetric density (23) in space.

 Einstein and Rosen first pointed out a difference of this kind in Section 1 of the fundamental work \cite{einstein}, having considered a simple one-dimensional boundary of a two-sheeted space. They showed that this one-dimensional boundary can be described both by a limiting transition to a thin material shell with a nonzero stress energy tensor and by an exact vacuum solution with a zero stress energy tensor. There are many examples in physics where the limiting transition and the exact solution give different results.

 \subsection{Regular black holes}

 The procedure of topological dressing makes it possible to transform a singular one-sheet state of a field and the space on which this field is defined into a two-sheeted state. At the same time, there is another way to eliminate the singularity. It is associated with the use, instead of the usual zero vacuum, for which the stress energy tensor is zero (\(T_{\mu \nu} = 0\)), of the isotropic de Sitter vacuum \(T_{\mu \nu} = \Lambda g_{\mu \nu} / (8 \pi)\), where \(\Lambda\) is the cosmological constant. In this case, Einstein's equations in vacuum have a spherically symmetric solution, free of singularities \cite{dymnikova,lan}. This solution, which has long-range asymptotics according to Schwarzschild and short-range asymptotics according to de Sitter, is called a regular black hole (RBH).

The RBH is indeed regular over the entire one-sheet space and, therefore, has finite energy. Therefore, applying the topological dressing procedure to it would not lead to the elimination of the singularity - it is already absent. However, such a procedure is also possible here and makes sense, since this turns on an additional option to reduce the energy.

Since RBH initially has two event horizons (the Schwarzschild horizon and the de Sitter horizon), the question of the traversability of a wormhole emerging from such an RBH requires a separate study.

\section{Conclusions}

Thus, using the covariance of the Einstein-Maxwell equations with respect to coordinate transformations, we constructed a method to transform an arbitrary one-sheet solution of these equations into a two-sheeted solution with a wormhole topology. The topological dressing method differs from the thin shell model \cite{visser,poisson,sharif,zubair} in that it gives an exact solution \textbf{ throughout the space, including the throat of the wormhole}.  This allows us to show that the throat of the wormhole itself is a source of gravitational and electromagnetic fields. The topological dressing method makes it unnecessary to search for a suitable stress-energy tensor and replaces it with an exact two-sheeted solution of the Einstein-Maxwell equations in vacuum. 

In the new approach, the centre of gravity shifts from the search for a suitable stress-energy tensor to the calculation of the action integral and the selection of conditions that ensure the stability of the already found wormhole. When applied to the massless one-sheet Reissner-Nordström solution, this leads to a two-sheeted solution describing a massless traversable wormhole.

Note that the gain in the action for the two-sheeted state (with a wormhole) compared to the one-sheet state (without a wormhole) consists in the effective exclusion from the space of a naked singularity, which gives a divergent positive contribution to the action. In the proposed topological dressing method, this exclusion is made by selecting a region \(\Omega\), the size and shape of which is determined by directly applying the principle of least action. This seems to mean that any one-sheet state with a naked singularity is unstable with respect to the formation of a two-sheeted wormhole.

This conclusion is quite general and should depend neither on the dimension of space nor on the specific form of the Lagrangian. Therefore, the topological dressing method can be used not only in classical general relativity but also in its various modifications.

\section*{Acknowledgements}

I am pleased to express my gratitude to Vadim Mogilevsky for discussions and comments.

\bibliography{main}

\end{document}